\documentclass[%
 aip,
 amsmath,amssymb,
 reprint,%
]{revtex4-1}

\usepackage{graphicx}
\usepackage{dcolumn}
\usepackage{bm}

\usepackage[utf8]{inputenc}
\usepackage[T1]{fontenc}
\usepackage{mathptmx}

\begin{document}

\preprint{AIP/123-QED}

\title{Symmetry in a space of conceptual variables}

\author{Inge S. Helland}
 \email{ingeh@math.uio.no}
 \affiliation{Department of Mathematics, University of Oslo \\ P.O.Box 1053 Blindern, N-0316 Oslo, Norway}

\date{\today}

\begin{abstract}

A conceptual variable is any variable defined by a person or by a group of persons. Such variables may be inaccessible, meaning that they cannot be measured with arbitrary accuracy on the physical system under consideration at any given time. An example may be the spin vector of a particle; another example may be the vector (position, momentum). In this paper, a space of inaccessible conceptual variables is defined, and group actions are defined on this space. Accessible functions are then defined on the same space. Assuming this structure, the basic Hilbert space structure of quantum theory is derived: Operators on a Hilbert space  corresponding to the accessible variables are introduced; when these operators have a discrete spectrum, a natural model reduction implies a new model in which the values of the accessible variables are the eigenvalues of the operator. The principle behind this model reduction demands that a group action may also be defined also on the accessible variables; this is possible if the corresponding functions are permissible, a term that is precisely defined. The following recent principle from statistics is assumed: every model reduction should be to an orbit or to a set of orbits of the group. From this derivation, a new interpretation of quantum theory is briefly discussed: I argue that a state vector may be interpreted as connected to a focused question posed to nature together with a definite answer to this question.  Further discussion of these topics is provided in a recent book published by the author of this paper.

\end{abstract}

\maketitle

\section{\label{sec:level1}Introduction}

The standard textbook treatment of quantum mechanics relies on a few axioms, of which two are the most basic. The first axiom assumes the Hilbert space structure. It states that to each physical variable there is a corresponding operator $A$, and that the possible values of that variable are the eigenvalues of $A$. The second axiom gives the probability structure, namely the Born formula. This paper will concentrate on the first axiom. I intend to show that it follows from a simple logical structure: a conceptual variable with a concrete group acting on the space over which this conceptual variable varies.

One purpose of the book [1] is to propose a common basis for statistical inference theory and quantum theory. To achieve this, it is crucial to take a definition of a conceptual variable as a point of departure. For this purpose a conceptual variable is any variable that can be defined in words by a person or by a group of persons in an experimental or observational situation, or more generally in any epistemic process. Behind this definition of conceptual variables there also lies an interpretation of quantum theory where the observer plays an essential role. Some conceptual variables are inaccessible, meaning that they cannot be assessed with full accuracy by any observer. An example can be  a spin or angular momentum vector of some particle. Another example is the vector (position, momentum) for a particle. However, some functions of inaccessible variables can be measured, and these functions are called e-variables (epistemic conceptual variables). Part of the philosophy of [1] is that the e-variables are closely related to the parameters of statistical inference; these parameters can be modelled as they are in statistical inference theory when data are present.

In the following discussion, I may also partly have in mind other situations than epistemic processes. In these cases a conceptual variable is any variable that can be defined by a person or by a group of communicating persons. Some such variables only have meaning for this specific group of people, and thus can not be generalized. Some do not have a precise meaning at all in their given context, in the sense that their components are mutually exclusive. These variables may tentatively be called inaccessible. The e-variables then correspond to less ambitious conceptual variables which can be precisely defined, at least to the group of people that define them. These notions are by necessity a bit vague. For more concrete ideas, look at the qualitative discussion of complementarity in Chapter 6 of [1]. Those who prefer so, may limit the discussion below to the epistemic process case.

However, in a more general context the notions of focusing and decisions are very important. These notions are used in statistical theory, but they should here be interpreted in their primitive sense.

It is crucial in our context here that we assume that there are defined transformation groups on the spaces of conceptual variables that are discussed. This may the group of all automorphisms on the space in which the conceptual variable varies, or it may be a concrete subgroup of this group.

In this paper I discuss such situations more closely with a focus on the more mathematical aspects of the notions and situations described above. I assume the existence of a concrete (physical) situation, and that there is a space $\Phi$ of the inaccessible conceptual variable $\phi$  with a group $K$ acting on this space. There is an e-variable, an accessible conceptual variable, $\theta$ defined, a function on $\Phi$. This $\theta$ varies on a space $\Theta$, and the group $K$ induces a transformation group $G$ on $\Theta$. A very simple situation is when $\phi$ is a spin vector, and $\theta$ is a spin component in a given direction, but this paper focuses on more general situations. When there are more potential e-variables, I will denote this by a superscript $a$: $\theta^a$ and $G^a =\{g^a\}$. Note that in Chapter 4 of [1] I used different notations for the groups: $G$ and $\tilde{G}^a$ for what is here called $K$ and $G^a$. The group $G^a$ which was refered to there, is here the subgroup of $K$ corresponding to $G^a$, denoted by $H^a$ below. Both here and in [1] I will use the word `group' as synonymous to `group action' or transformation group on some set, not as an abstract group.

\section{Basic setting}

In Chapter 3 of [2] right-hand notation is used for group transformations: $\phi\mapsto \phi k$ for $k\in K$ and $\theta\mapsto \theta g$ for $g\in G$. This allows right-invariant measures to be defined in a simple way and has also the following simple advantage: If $g_1$ is applied first to $\Theta$ followed by $g_2$, then the composed transformation is $g_1 g_2$, rather than $g_2 g_1$. However, in this paper I will need the left-invariant measure, so I will use the more familiar notation: $\phi\mapsto k\phi [=k(\phi)]$ and $\theta\mapsto g\theta$.

Starting with $\Phi$ and the group $K$ acting on $\Phi$, let $\theta(\cdot)$ be a function on $\Phi$, and let $\Theta$ be the range of this function.

As discussed in the Introduction, I regard `accessible' and `inaccessible' as primitive notions. $\Theta$ and $\Phi$ are equipped with topologies, and all functions are assumed to be Borel-measurable.
\bigskip

 \textit{Definition 1. 
The e-variable $\theta$ is maximally accessible if the following holds: If $\theta$ can be written as $\theta=f(\psi)$ for a function $f$ that is not one-to-one, the conceptual variable $\psi$ is not accessible. In other words:$\theta$ is maximal under the partial ordering defined by $\alpha\le \beta$ iff $\alpha=f(\beta )$ for some function $f$.}
\bigskip

Note that this partial ordering is consistent with accessibility: If $\beta$ is accessible and $\alpha=f(\beta )$, then $\alpha$ is accessible. Also, $\phi$ is an upper bound under this partial ordering. The existence of maximal accessible conceptual variables follows then from Zorn's lemma.
\bigskip

\textit{Definition 2.
The e-variable $\theta$ is called\emph{ permissible} if the following holds: $\theta(\phi_1)=\theta(\phi_2)$ implies $\theta(k\phi_1 )=\theta(k\phi_2 )$ for all $k\in K$.}
\bigskip

With respect to parameters and subparameters along with their estimation, the concept of permissibility is discussed in some details in Chapter 3 in [2]. The main conclusion, which also can be generalized to this setting, is that under the assumption of permissibility one can define a group $G$ of actions on $\Theta$ such that

\begin{equation}
(g\theta)(\phi):=\theta(k\phi);\ k\in K.
\label{kg}
\end{equation}

Herein I use different notations for the group actions $g$ on $\Theta$ and the group actions $k$ on $\Phi$; in contrast,  the same symbol $g$ was used in [2]. The background for that is
\bigskip

\textit{Lemma 1.
Assume that $\theta$ is a permissible e-variable. The function from $K$ to $G$ defined by (\ref{kg}) is then a group homomorphism.}
\bigskip

\textit{Proof.} Let $k_i$ be mapped upon $g_i$ by (\ref{kg}) for $i=1,2$. Then, for all $\phi\in \Phi$ we have $(g_i\theta )(\phi)=\theta(k_i\phi )$. Assume that $k_2\phi $ is mapped to $\theta'=\theta(k_2 \phi )=( g_2\theta)(\phi)$. Then also $\theta(k_1 k_2 \phi  )=(g_1\theta')(k_2\phi )=(g\theta)(\phi)$ for some $g$. Thus $(g_1(g_2\theta))(\phi)=(g\theta)(\phi)$ for all $\phi$, and since the mapping is permissible, we must have $g=g_1g_2$.
\bigskip

In the statistical context of parameters and subparameters, the concept of permissibility is connected to several deep problems, as briefly discussed in [2]. First, one should mention Peter McCullagh's general categorical theory-based requirements for statistical models, see [3]. Next, under transitivity and other week assumptions it implies that Bayesian credibility sets based on a right invariant prior and frequentist confidence sets are numerically equal. Finally, the assumption of permissibility and the use of a right invariant prior turn out to be enough to eliminate the marginalization paradoxes of Dawid et al. [4], along with some similar inconsistencies, see [5].

In the more general context considered here, it is also important to define left and right invariant measures, both on the groups and on the spaces of conceptual variables. In the mathematical literature, see for instance [6,7], Haar measures on the groups are defined (assuming locally compact groups). Right ($\mu_K$) and left ($\nu_K$) Haar measures on the group $K$ satisfy
\begin{eqnarray*}
\mu_K(Dk)=\mu_K(D), \ \mathrm{and}\ \nu_K(kD)=\nu_K(D)\\ \mathrm{for}\ k\in K\ \mathrm{and}\ D\subset K,\ \mathrm{respectively}.
\end{eqnarray*}

Next define the corresponding measures on $\Phi$. As is commonly done, I assume that the group operations $(k_1,k_2)\mapsto k_1k_2$, $(k_1,k_2)\mapsto k_2k_1$ and $k\mapsto k^{-1}$ are continuous. Furthermore, I will assume that the action $(k,\phi)\mapsto k\phi$ is continuous. 

As discussed in Wijsman [8], an additional condition is that every inverse image of compact sets under the function $(k,\phi)\mapsto (k\phi,\phi)$ should be compact. A continuous action by a group $K$ on a space $\Phi$ satisfying this condition is called \emph{proper}. This technical condition turns out to have useful properties and is assumed throughout this paper. When the group action is proper, the orbits of the group can be proved to be closed sets relative to the topology of $\Phi$.

The connection between $\nu_K$ defined on $K$ and the corresponding left invariant measure $\nu$ defined on $\Phi$ is relatively simple: If for some fixed value $\phi_0$ of the conceptual variable the function $\beta$ on $K$ is defined by $\beta: k\mapsto k\phi_0$, then $\nu(E)=\nu_K (\beta^{-1}(E))$.This connection between $\nu_K$ and $\nu$ can also be written $\nu_K(dk)=d\nu(k\phi_0))$, so that $d\nu(hk\phi_0)=d\nu (k\phi_0)$ for all $h, k \in K$.

The following result, originally due to Weil is proved in [8]; for more details on the right invariant case, see also [2].
\bigskip

\textbf{Theorem 1.}
\textit{The left-invariant measure measure $\nu$ on $\Phi$ exists if the action of $K$ on $\Phi$ is proper and the group is locally compact.}
\bigskip

Note that $\nu$ can be seen as an induced measure on each orbit of $K$ on $\Phi$, and it can be arbitrarily normalized on each orbit. $\nu$ is finite on a given orbit if and only if the orbit is compact. In particular, $\nu$ can be defined as a probability measure on $\Phi$ if and only if all orbits of $\Phi$ are compact. Furthermore, $\nu$ is unique only if the group action is transitive.

In a corresponding fashion, a right invariant measure can be defined on $\Phi$. This measure satisfies $d\mu (kh\phi_0)=d\mu (k\phi_0)$ for all $k,h\in K$. In many cases the left invariant measure and the right invariant measure are equal.

\section{Operators and quantization}

In the quantum-mechanical context defined in [1], $\theta=\theta(\phi)$ is an accessible e-variable, and one should be able to introduce an operator associated with $\theta$. The following discussion which is partly inspired by [9]. considers an irreducible unitary representation of $K$ on a complex Hilbert space $\mathcal{K}$.

\subsection{Brief discussion of group representation theory}
 \label{sec:3.1}
 
A group representation of $K$ is a continuous homomorphism from $K$ to the group of invertible linear operators $V$ on some vector space $\mathcal{K}$:
\begin{equation}
V(k_1 k_2 )=V(k_1 )V(k_2 ).
\label{1}
\end{equation}
It is also required that $V(e)=I$, where $I$ is the identity, and $e$ is the unit element of $K$. This assures that the inverse exists: $V(k)^{-1}=V(k^{-1})$. The representation is unitary if the 
operators are unitary ($V(k)^{\dagger}V(k)=I$). If the vector space is finite-dimensional, we have a representation $D(V)$ on the square, invertible matrices. For any 
representation $V$ and any fixed invertible operator $U$ on the vector space, we can define a new equivalent  representation as $W(k)=UV(k)U^{-1}$. One can prove that two 
equivalent unitary representations are unitarily equivalent; thus $U$ can be chosen as a unitary operator.

A subspace $\mathcal{K}_1$ of $\mathcal{K}$ is called invariant with respect to the representation $V$ if $u\in \mathcal{K}_1$ implies $V(k)u\in \mathcal{K}_1$ for all $k\in K$. The null-space $\{0\}$ and the whole space
$\mathcal{K}$ are trivially invariant; other invariant subspaces are called proper. A group representation $V$ of a group $K$ in $\mathcal{K}$ is called irreducible if it has no proper invariant subspace.
A representation is said to be fully reducible if it can be expressed as a direct sum of irreducible subrepresentations. A finite-dimensional unitary representation of any group 
is fully reducible. In terms of a matrix representation, this means that we can always find a  $W(k)=UV(k)U^{-1}$ such that $D(W)$ is of minimal block diagonal form. Each one of 
these blocks represents an irreducible representation, and they are all one-dimensional if and only if $K$ is Abelian. The blocks may be seen as operators on subspaces of the 
original vector space, i.e., the irreducible subspaces. The blocks are important in studying the structure of the group.

A useful result is Schur's Lemma, (see for instance [10]):
\bigskip

\textit{Let $V_1$ and $V_2$ be two irreducible representations of a group $K$; $V_1$ on the space $\mathcal{K}_1$ and $V_2$ on the space $\mathcal{K}_2$. Suppose that there exists a linear map $T$ from $\mathcal{K}_1$ to 
$\mathcal{K}_2$ such that}
\begin{equation}
V_2 (k)T(v)=T(V_1 (k)v)
\label{2}
\end{equation}
\textit{for all $k\in K$ and $v\in\mathcal{K}_1$.}

\textit{Then either $T$ is zero or it is a linear isomorphism. Furthermore, if $\mathcal{K}_1=\mathcal{K}_2$, then $T=\lambda I$ for some complex number $\lambda$.}
\bigskip

Let $\nu$ be the left-invariant measure of the space $\Phi$ induced by the group $K$, and consider the Hilbert space $\mathcal{K}=L^2 (\Phi ,\nu)$. Then the left-regular 
representation of $K$ on $\mathcal{K}$ is defined by $U^{L}(k)f(\phi)=f(k^{-1}\phi)$. This representation always exists, and it can be shown to be unitary, see [10].

If $V$ is an arbitrary representation of a compact group $K$ in $\mathcal{K}$, then there exists in $\mathcal{K}$ a new scalar product defining a norm equivalent to the initial one, relative to which $V$ 
is a unitary representation of $K$.

For references to some of the vast literature on group representation theory, see Appendix A.2.4 in [2].

\subsection{A resolution of the identity}
\label{sec:3.2}

In the following I assume that the group $K$ (and later also the groups $G$ and $H$) have representations that give square-integrable coherent state systems (see page 43 of [11]). For instance this is the case for all representations of compact semisimple groups, representations of discrete series for real semisimple groups, and some representations of solvable Lie groups.

Let $K$ be an arbitrary such group, and let $V(k)$ be one of its unitary  irreducible representations acting on the Hilbert space $\mathcal{K}$. Assume that $K$ is acting transitively on a space $\Phi$, and fix $\phi_0\in\Phi$. Then every $\phi\in\Phi$ can be written as $\phi=k\phi_0 $ for some $k\in K$.

Also, fix a vector $|\phi_0\rangle\in\mathcal{K}$, and define the coherent states $|\phi\rangle=|\phi(k)\rangle=V(k)|\phi_0\rangle$. With $\nu$ being the left invariant measure on $\Phi$, introduce the operator
\begin{equation}
T=\int |\phi(k)\rangle\langle\phi(k)|d\nu(k\phi_0).
\label{3}
\end{equation}
Note that the measure here is over $\Phi$, but the elements are parametrized by $K$.
\bigskip

\textit{Lemma 2.
$T$ commutes with every $V(h); h\in K$.} 
\bigskip

\textit{Proof.} $\ \ V(h)T=$
\begin{eqnarray*}
\int V(h) |\phi(k)\rangle\langle\phi(k)|d\nu(\phi_0 k)
=\int |\phi(hk)\rangle\langle\phi(k)|d\nu(k\phi_0)\\=\int |\phi(r)\rangle\langle\phi(h^{-1}r)|d\nu(h^{-1}r\phi_0 ).
\end{eqnarray*}
Since $|\phi(h^{-1}r)\rangle=V(h^{-1}r)|\phi_0\rangle =V(h^{-1})V(r)|\phi_0\rangle =V(h)^\dagger |\phi(r)\rangle$, we have $\langle \phi(h^{-1}r)|=\langle\phi(r)|V(h)$, and since the measure $\nu$ is left-invariant, it follows that $V(h)T=TV(h)$.
\bigskip

From the above and Schur's Lemma it follows that $T=\lambda I$ for some $\lambda$. Since $T$ by construction only can have postive eigenvalues, we must have $\lambda >0$. Defining the measure $d\mu(\phi)=\lambda^{-1}d\nu(\phi)$ we therefore have
\begin{equation}
\int |\phi\rangle\langle\phi |d\mu(\phi)=I.
\label{4}
\end{equation}
For a more elaborate similar construction taking into account the socalled isotropy subgroup, see Chapter 2 of [11].

\subsection{Focusing and a new resolution of the identity}
\label{sec:3.3}

Again let $\Phi$ be the space of inaccessible conceptual variables, and let us focus on a permissible accessible function $\theta (\cdot )$ on $\Phi$. Then, using (\ref{kg}) we can define a group $G$ acting upon the range $\Theta$ of $\theta(\cdot)$. In principle $\theta(\cdot)$ and thus $G$ can be quite arbitrary. However I will now introduce a recent principle from statistical inference. For  discussion, examples and motivation, see Section 2.2 in [1].
\smallskip

\textit{Principle. Every model reduction of a statistical model should be to an orbit or to a set of orbits of the group when a group is defined on the parameter space.} 
\smallskip

I will now extend this principle to the quantum setting, consider a model reduction where $\Theta$ is reduced to a subset, assuming that this model reduction is to a simple orbit of the group $G$. But then $G$ is transitive on the new $\Theta$, and the construction of the previous section can be repeated.

Explicitly, fix $\theta_0\in\Theta$ (the reduced space). Then every $\theta\in\Theta$ can be written as $\theta=g\theta_0$ for some $g\in G$. Also, let $U(g)$ be a unitary irreducible representation of $G$ acting on a Hilbert space $\mathcal{H}$, and fix a vector $|\theta_0\rangle\in\mathcal{H}$. Define the coherent states $|\theta\rangle=U(g)|\theta_0\rangle$ when $\theta=g\theta_0$. Then by the argument of the previous section there exists a measure $\rho$ on $\Theta$ such that
\begin{equation}
\int |\theta\rangle\langle\theta| d\rho(\theta)=I.
\label{5}
\end{equation}
For later use it may be convenient to treat $\theta$ explicitly as a function on $\Phi$, and let $\rho$ be the marginalization of a measure $\tau$ on $\Phi$. Then it follows that
\begin{equation}
\int |\theta(\phi)\rangle\langle\theta(\phi)|d\tau(\phi)=I.
\label{6}
\end{equation}

Note that if $|\psi(\phi)\rangle=V|\theta(\phi)\rangle $ for some unitary operator $V$, then an equivalent resolution of the identity is
\begin{equation}
\int |\psi(\phi)\rangle\langle\psi(\phi)|d\tau(\phi)=I.
\label{7}
\end{equation}

\subsection{Quantum operators}
\label{sec:3.4}

Again let $\Phi$ be the space of inaccessible conceptual variables, and let $\theta(\phi)$ be an accessible e-variable that is assumed to be permissible with respect to the group $K$.

In general, an operator corresponding to $\theta$ may be defined by
\begin{eqnarray}
A=A^\theta=\int \theta(\phi)|\theta(\phi)\rangle\langle\theta(\phi) |d\tau(\phi)\\
=\int \theta |\theta\rangle\langle\theta | d\rho (\theta).
\label{8}
\end{eqnarray}
$A$ is defined on a domain $D(A)$ of vectors $|v\rangle\in\mathcal{H}$ where the integral defining $\langle v|A|v\rangle$ converges.

This mapping from an e-variable $\theta$ to an operator $A$ has the following properties:

(i) If $\theta=1$, then $A=I$.

(ii) If $\theta$ is real-valued, then $A$ is symmetric (for a definition of this concept for operators and its relationship to self-adjointness, see [12].)

(iii) The change of basis through a unitary transformation is straightforward.

For further important properties, we need some more theory.
\bigskip

\textbf{Theorem 2.} \textit{For the reduced model, let $H$ be the subgroup of $K$ consisting of any transformation $h$ such that $\theta(h\phi)=g\theta(\phi)$ for some $g\in G$. Then $H$ is the maximal group under which the e-variable $\theta$ is permissible.}
\bigskip

\textit{Proof.} Let $\theta(\phi_1) =\theta(\phi_2)$ for all $\theta\in\Theta$. Then for $h\in H$ we have $\theta(h\phi_1)=g\theta(\phi_1)=g\theta(\phi_2)=\theta (h\phi_2)$, thus $\theta$ is permissible under the group $H$. For a larger group, this argument does not hold.
\bigskip

 For $h\in H$ we have a mapping $h\rightarrow g$ for some $g\in G$. For this $g$ define $V(h)=U(g)$. Then it is easy to see that $V(h)$ is a unitary representation of $H$ on some subspace of $\mathcal{K}$. 
\bigskip

\textbf{Theorem 3.} \textit{ For $h\in H$, $V(h^{-1})AV(h)$ is mapped by $\theta'(\phi)=\theta(h\phi)$.}
\bigskip

\textit{Proof.} $V(h^{-1})AV(h)=$
\begin{eqnarray*}
\int \theta(\phi)|\theta(h^{-1}\phi)\rangle\langle\theta(h^{-1}\phi) |d\tau(h^{-1}\phi)\\
=\int \theta(h\phi)|\theta(\phi)\rangle\langle\theta(\phi) |d\tau(\phi)
\end{eqnarray*}
Here, $V(h)$ is a unitary representation, and $\rho$ can be taken as a left-invariant measure (cf. the construction in subsection B).

An equivalent conclusion is that $U(g^{-1})AU(g)$ is mapped by $\theta'=g\theta$ for $g\in G$.
\bigskip

Further properties of the mapping from $\theta$ to $A$ may be developed in a similar way.
The mapping corresponds to the usual way that the operators are allocated to observables in the quantum mechanical literature. But note that this mapping comes naturally here from the notions of conceptual variable and e-variables on which group actions are defined.

For any Borel-measurable function $f$ of $\theta$, one can define an operator corresponding to (\ref{8}):
\begin{equation}
A^{f(\theta)}=\int f(\theta)|\theta\rangle\langle\theta |d\rho (\theta).
\label{9}
\end{equation}
Important special cases include $f(\theta)=I(\theta\in B)$ for sets $B$, which can be related to the spectral theorem. 

Another important case is connected to inference theory in the way it is advocated in [1]. Assume that there are data $z$ and a statistical model for these data of the form $P(z\in C|\theta)$ for sets $C$. Then a positive operator-valued measure (POVM) on the data space can be defined by
\begin{equation}
M(C)=\int P(z\in C |\theta)|\theta\rangle\langle\theta |d\rho (\theta).
\label{10}
\end{equation}
The density of $M$ at a point $z$ is called the likelihood effect in [1], and is the basis for the focused likelihood principle formulated there.

Finally, given a probability measure with density $\pi(\theta)$ over the values of $\theta$, one can define a density operator $\sigma$ by
\begin{equation}
\sigma=\int \pi(\theta)|\theta\rangle\langle\theta |d\rho (\theta).
\label{11}
\end{equation}

In [1] the probability measure $\pi$ was assumed to have one out of three possible interpretations: 1) as a Bayesian prior,  2) as a Bayesian posterior or 3) as a frequentist confidence distribution (see [13]).

Assume that $\theta$ is real-valued. Then based on the spectral theorem (e.g., [12]) we have that there exists a projectionvalued measure, $E$ on $\Theta$ such that for $|v\rangle\in D(A)$
\begin{equation}
\langle v|A|v\rangle =\int_{\sigma(A)} \theta d\langle v|E(\theta)|v\rangle .
\label{12}
\end{equation}
Here $\sigma(A)$ is the spectrum of $A$ as defined in [12].
The case with a discrete spectrum is discussed in the next subsection.

\subsection{The model reduction as a quantization}
\label{sec:3.5}

Note that the construction in subsection D can be made for any conceptual variable $\phi$ and for any e-variable $\theta=\theta(\phi)$. I will assume that $A$ has a purely discrete spectrum. Let the eigenvalues be $\{u_j\}$ and let the corresponding eigenspaces be $\{V_j\}$. The vectors of these eigenspaces are defined as quantum states, and as in [1], each eigenspace $V_j$ is associated with a question `What is the value of $\theta$?' together with a definite answer `$\theta =u_j$'. This assumes that the set of values of $\theta$ can be reduced to this set of eigenvalues, which I will attempt to justify as follows.

Recall the model reduction principle formulated in the beginning of subsection C.
\bigskip

\textbf{Theorem 4.}
\textit{Assume that the e-variable $\theta$ is permissible, which allows the group $G$ on $\Theta$ to be defined. Let $\{u_j\}$ be the eigenvalues of the operator $A$ corresponding to $\theta$. Then $\cup_j \{\theta:\theta=u_j\}$ is an orbit or several orbits under the group $G$. If the set of eigenvalues constitute just one orbit, then $G$ is the permutation group on these eigenvalues.}
\bigskip

\textit{Proof.} For each $j$, let $|j\rangle$ be an eigenvector of $A$ with eigenvalue $u_j$, and let $g\in G$. Then $g\theta(\phi )=\theta(h\phi)$ for at least some $h\in H$. By Theorem 3 we have that the operator $V(h^{-1})AV(h)$ is mapped by $g\theta(\phi )$. Assume now that $\theta_0 =u_j$ for some $j$. We need to show that $g\theta_0$ is another eigenvalue for $A$, which follows from the fact that $|V(h^{-1})AV(h)-\lambda I|=|A-\lambda I|$, so that these two determinants have the same zeros.

Let $I_0 =\{u_j : u_j =g\theta_0\ \mathrm{for\ some}\ g\in G\}$. Then this is an orbit for $G$. Either the orbit contains all eigenvalues or there is an eigenvalue $\theta_1$ not in $I_0$. Then let $I_1 =\{u_j : u_j =g\theta_1\ \mathrm{for\ some}\ g\in G\}$ and continue. Sooner or later all eigenvalues are reached, and we obtain the set of eigenvalues as a union of orbits of $G$.
\bigskip

In subsection D it was assumed that $G$ in the final reduced model was transitive. Then the set of eigenvalues constitute just one orbit. \emph{The model reduction determined by this case constitutes a natural quantization procedure.}
\bigskip

We also have the following:
\bigskip

\textbf{Theorem 5.} \textit{Assume that the set of measurable values of $\theta$ is restricted to the above eigenvalues. Then $\theta$ is maximally accessible if and only if each eigenspace $V_j$ is one-dimensional.}
\bigskip

\textit{Proof.} The assertion that there exists an eigenspace that is not one-dimensional, is equivalent with the following: Some eigenvalue $u_j$ correspond to at least two orthogonal eigenvectors $|j\rangle$ and $|i\rangle$. Based on the spectral theorem, the operator $A$ corresponding to $\theta$ can be written as $\sum_r u_r P_r$, where $P_r$ is the projection upon the eigenspace $V_r$. Now define a new e-variable $\psi$ whose operator $B$ has the following properties: If $r\ne j$, the eigenvalues and eigenspaces of $B$ are equal to those of $A$. If $r=j$, $B$ has two different eigenvalues on the two one-dimensional spaces spanned by $|j\rangle$ and $|i\rangle$, respectively, otherwise its eventual eigenvalues are equal to $u_j$ in the space $V_j$. Then $\theta=\theta(\psi)$, and $\psi\ne\theta$ is inaccessible if and only if $\theta$ is maximally accessible. This construction is impossible if and only if all eigenspaces are one-dimensional.

\section{Coupling different focusings together}
 \label{sec:4.}
 
 \subsection{The maximal case}
 
In this section consider a Hilbert space $\mathcal{H}$ of finite dimension $n$. Again let $\phi$ be an inaccessible conceptual variable. For an index set $\mathcal{A}$, focus on $\lambda^a$ for $a\in\mathcal{A}$, a set of maximally accessible e-variables. Note that each $\lambda^a$ corresponds to a unique operator $A^a$, and that this operator has the spectral decomposition
\begin{equation}
A^a=\sum_j u_j^a |a;j\rangle\langle a;j|.
\label{13}
\end{equation}

\smallskip
By maximality, only one $\lambda^a$ can be measured on the system at a given time. This is a manifestation of Niels Bohr's complementarity.

The following is proven in [1] under certain technical conditions, and also in the case of spin/ angular momentum: given a vector $|v\rangle \in \mathcal{H}$, there is at most one pair $(a,j)$ such that $|a;j\rangle=|v\rangle$. The main interpretation in [1] is motivated as follows: Suppose the existence of such a vector $|v\rangle$ with $|v\rangle=|a;j\rangle$ for some $a$ and $j$. Then the fact that the state of the system is $|v\rangle$ means that one has focused on a question (`What is the value of $\lambda^a$?') and obtained the definite answer ($\lambda^a=u_j^a$.) The question can be associated with the orthonormal basis  $\{|a;j\rangle ;j=1,2,...,n\}$. 

After this we are left with the problem of determining conditions under which \emph{all} vectors $|v\rangle\in \mathcal{H}$ can be interpreted as above. This will require a richn index set $\mathcal{A}$. This problem will not be considered further here. However, it should be noted, this richness requirement holds in the simple case of a spin 1/2 particle, as discussed in [1]. 

The following simple observation should also be noted: Trivially, every vector $|v\rangle$ is the eigenvector of \emph{some} operators. Assume that there is one such operator $A$ that is physically meaningful, and for which $|v\rangle$ is also a non-degenerate eigenvector. Let $\lambda$ be a physical variable associated with $A$. Then $|v\rangle$ can again be interpreted as a question (`What is the value of $\lambda$?') along with a definite answer to this question.

\subsection{The general case}

First go back to the maximal symmetrical epistemic setting. Let again $\lambda^a=\lambda^a(\phi)$ be as in the previous subsection. Let $t^a$ be an arbitrary function on the range of $\lambda^a$, and let us focus on $\theta^a =t^a(\lambda^a)$ for each $a\in\mathcal{A}$.

 Let the Hilbert space be as in the previous subsection, and suppose that it has an orthonormal basis that can be written in the form $|a;i\rangle$ for $i=1,...,n$. Let $\{u_i^a\}$ be the values of $\lambda^a$, and let $\{s_j^a\}$ be the values of $\theta^a$. Define $C_j^a=\{i:t^a(u_i^a)=s_j^a\}$, and let $V_j^a$ be the space spanned by $\{|a;i\rangle :i\in C_j^a\}$. Let $\Pi_j^a$ be the projection upon $V_j^a$.
 
 Then we have the following interpretation of any $|a;i\rangle\in V_j^a$. (1) the question: `What is the value of $\theta^a$?' has been posed, and (2) we have obtained the answer $\theta^a=s_j^a$. Note that in this case, several pairs $(a,i)$ correspond to a given vector $|v\rangle$.
 
 From the above construction we may also define the operator connected to the e-variable $\theta^a$ as
 \begin{equation}
 A^a=\sum_j s_j^a \Pi_j^a =\sum_i t^a(u_i^a)|a;i\rangle\langle a;i|.
 \label{14}
\end{equation}
 Note that this gives all possible states and all possible values corresponding to the accessible e-variable $\theta^a$. Unless the function $t^a$ is one-to-one, the operator $A^a$ has no longer distinct eigenvalues. 

 \section{Simple examples}
 
 \subsection{Spin/ angular momentum and the rotation group}
 
 The group theory applied in this and in the following examples is well known and discussed in depth in many books. Here, the point is to consider the group as a group action on a space of conceptual variables. To this end, consider a spin or angular momentum vector $\phi$ with fixed norm $r$ varying on a sphere $\Phi$ with radius $r$. This vector is inaccessible. However, given some direction $a$, the components $\theta^a =a\cdot \phi$ can be measured and are e-variables. After model reduction/quantization, each $\theta^a$ takes the values $-j,-j+1,...,j-1,j$ for some integer or half-integer $j$.
 
 The group $K$ consists of rotations of the vector $\phi$, while $G^a$ is simply the permutation group on the values of $\theta^a$. The (left) invariant measure for $K$ is the natural symmetry measure on the sphere.
 
 The operator $A^a$ can be written $A^a =a\cdot J$ for some vector operator $J=(J_x, J_y, J_z )$ acting on a Hilbert space $\mathcal{H}$ of dimension $2j+1$. It is well known (see for example [13]), that by taking $\hbar=1$, the irreducible representations of the group $K$ can be written as
 \begin{equation}
 V(k)=\mathrm{exp} (-i\phi_k \cdot J),
 \label{15}
\end{equation}
 where the rotation $k$ is represented by the rotation angle $\phi_k$.
 
 \subsection{Position and momentum}
 
 Consider the one-dimensional case, where a point in the phase space is given by $\phi=(\xi,\pi)$,  $\xi\in\Xi$ and $\pi\in\Pi$ are the position and the momentum of some particle, respectively. The vector $\phi$ is inaccessible, while both $\xi$ and $\pi$ are accessible e-variables. 
 
 We can let the group $K$ consist of simultaneous translations of $\xi$ and $\pi$:
 \begin{equation}
 \xi\rightarrow \xi+c,\ \ \pi\rightarrow \pi+d.
 \label{16}
\end{equation}
 The invariant measure for this group is $d\nu=d\xi d\pi$, and the groups $G^\xi$ and $G^\pi$ consist of separate transitions. As discussed in [1] and elsewhere, a Hilbert space can be chosen by a position representation or by a momentum representation. In the first case, we have $\mathcal{H}=\mathrm{L^2}(\Xi,d\xi)$, the operator $A^\xi$ consists of multiplication with $\xi$, and ($\hbar=1$ again) $A^\pi$ is a differentiation operator $i^{-1}d/d\xi$, both operators with suitable domains of definitions. These operators have continuous spectra; no model reduction is called for.
 
 \section{Concluding remarks}
 
 Group theory and quantum mechanics are intimately connected, as discussed in details in [14] for example. The purpose of this article is to show that the familiar Hilbert space formulation can be derived mathematically from a simple basis of groups acting on conceptual variables. The consequences of this is further discussed in [1]. This discussion also provides a link to statistical inference, a link that will be further discussed elsewhere.
 
 However, we already here must view the existence of such a connection between quantum theory and statistical inference theory as an essential assumption. From the viewpoint of purely statistical inference the e-variables discussed in this paper are parameters. In many such situations also, it is useful to have a group of actions $G$ defined on the parameter space; see for instance the discussion in [15]. In the present paper, the quantization of quantum mechanics is derived from the following principle: all model reductions in some given model should be to an orbit or to a set of orbits of the group $G$.
 
 It is of some interest that the same criterion can be used to derive the statistical model corresponding to the partial least squares algorithm in chemometrics [16], and also to motivate the more general recently proposed envelope model [17].
 
This paper focuses on group symmetry in spaces of conceptual variables, partucularly those referred to as  parameters/e-variables in [1]. When data are present, in statistical inference one often starts with a group $G^*$ on the data space (see for instance [18]). Referring to the statistical model $P^\theta (X\in B)$ for Borel-sets $B$ in the data-space, the relationship is given by $P^{g\theta}(X\in B)=P^\theta (X\in g^* B)$. This induces a group homomorphism from $G^*$ to $G$. In some cases, this is an isomorphism, and the same symbol $g$ is often used in the data space and the parameter space. However, this is \emph{not} the case when the data space is discrete and the parameter space is continuous, as when the model is given by a binomial distribution or a Poisson distribution. This case is studied in detail in [19]. In these cases, group-theoretical methods were first used to construct the Poisson family and the binomial family, and the basic tool is coherent states for certain groups (the Weyl-Heisenberg group in the Poisson case and the group SU(2) in the binomial case). Finally, inference is studied by reversing the roles of data and parameter. The result in both cases is equivalent to Bayesian inference with a uniform prior on the parameter, which may be a coincidence. Taking this and the present paper as a point of departure, there seems to be a possibility to provide new ideas to symmetry-based approaches to statistical inference, and perhaps a possibility to relate this to quantum measurement theory; however, this problem area remains to be addressed in the future.
 
 As an extension of [19], a large class of probability distributions are shown to have connections to coherent states in [20]. For a general reference regarding coherent states, see [21].
 
 In the present paper, the first axiom of quantum theory is derived from reasonable assumptions. The second axiom, the Born formula, is derived in [1] based on the following: 1) a focused version of the likelihood principle from statistical inference, and 2) an assumption of perfect rationality, as expressed by the Dutch book principle.
 
 As also briefly stated in [1], one can perhaps expect after this, that such a relatively simple conceptual basis for quantum theory may facilitate a further discussion regarding its relationship to relativity theory. However, such considerations go far beyond the scopes of both [1] and the present paper. 
 
 \section*{Acknowledgments}
 
 I am grateful to Bj\o rn Solheim and to an anonymous referee for comments that resulted in an essential improvement to this paper.

\section*{References}

\setlength\parindent{0cm}

[1] Helland, I.S. ``Epistemic Processes. A Basis for Statistics and Quantum Theory.'' Springer, Berlin.(2018)

[2] Helland, I.S. ``Steps Towards a Unified Basis for Scientific Models and Methods.'' World Scientific, Singapore.(2010).

[3] McCullagh, P.  ``What is a statistical model?'' Annals of Statistics 30, 1225-1310.(2002).

[4] Dawid, A.P., Stone, M. and Zidek, J.V. ``Marginalization paradoxes in Bayesian and structural inference.'' J. Royal Statistical Society B  35, 189-233. (1973).

[5] Helland, I.S. ``Discussion of McCullagh, P. What is a statistical model?'' Annals of Statistics 30, 1225-1310. (2002).

[6] Nachbin, L. ``The Haar Integral.'' Van Nostrand, Princeton, NJ. (1965)

[7] Hewitt, E. and Ross, K.A. ``Abstract Harmonic Analysis, II.'' Springer-Verlag, Berlin. (1970).

[8] Wijsman, R.A. ``Invariant Measures on Groups and Their Use in Statistics.''  Lecture Notes - Monograph Series 14, Institute of Mathematical Statistics, Hayward, California. (1990).

[9] Bargeron, H., Curado, E.M.F., Gazeau, J.-P. and Rodrigues, L.M.C.S.  ``A baby Majorana quantum formalism.'' arXiv: 1701.0426 (2018).

[10] Barut, A.S. and Raczka, R. ``Theory of Group Representation and Applications.'' Polish Scientific 
Publishers, Warsaw. (1985).

[11] Perelomov, A.  ``Generalized Coherent States and Their Applications.'' Springer-Verlag, Berlin. (1986).

[12] Schweder, T. and Hjort, N.L. ``Confidence, Likelihood, Probability. Statistical Inference with Confidence Distributions.'' Cambridge University Press. (2016).

[13] Hall, B.C. ``Quantum Theory for Mathematicians.'' Graduate Texts in Mathematics, 267, Springer, Berlin. (2013).

[14] Greiner, W. and M\"{u}ller, B. ``Quantum Mechanics. Symmetries.'' Springer, Berlin. (1994).

[15] Helland, I.S. ``Statistical inference under symmetry.'' International Statistical Review 72, 409-422. (2004).

[16] Helland, I.S., S\ae b\o, S. and Tjelmeland, H. ``Near optimal prediction from relevant components.'' Scandinavian Journal of Statistics 39, 695-713. (2012).

[17] Helland, I.S., S\ae b\o, S., Alm\o y, T. and Rimal, R. ``Model and estimators for partial least squares.'' Journal of Chemometrics 32:  e3044. (2018).

[18] Helland, I.S. ``Statistical inference under symmetry.'' International Statistical Review 72, 409-422. (2004).

[19] Heller, B. and Wang, M. ``Group invariant inferred distributions via noncommutative probability.'' Recent Developments in Nonparametric Inference and Probability, IMS Lecture Notes - Monograph Series 50, 1-19. (2006).

[20] Ali, S.T. Gazeau, J.-P. and Heller, B. ``Coherent states and Bayesian duality.'' J. Phys. A: Math. Theor. 41, 365302. (2008).

[21] Gazeau, J.-P. ``Coherent States in Quantum Physics.'' Wiley, Weinheim. (2009).

\end{document}